# Transparent and flexible high-power supercapacitor based on carbon nanotube fibre aerogels†

Evgeny Senokos[a,b], Moumita Rana[a], Maria Vila[a], Julio Fernández-Cestau[a], Rubén D. Costa[a], Rebeca Marcilla[b,*] and Juan Jose Vilatela[a,*]

In this work, we report on the fabrication of continuous transparent and flexible supercapacitors by depositing a CNT network onto a polymer electrolyte membrane directly from an aerogel of ultra-long CNTs produced floating in the gas phase. The supercapacitors combine record power density of 1370 kW kg$^{-1}$ at high transmittance (*ca.* 70%), high electrochemical stability during extended cycling (>94% capacitance retention over 20,000 cycles) as well as against repeated 180° flexural deformation. They represent a significant enhancement of 1-3 orders of magnitude compared to the prior-art transparent supercapacitors based on graphene, CNTs, and rGO. These features mainly arise from the exceptionally long length of the CNTs, which makes the material behave as a bulk conductor instead of an aspect ratio-limited percolating network, even for electrodes with >90% transparency. The electrical and capacitive figures-of-merit for the transparent conductor are FoM$_e$ = 2.7, and FoM$_c$ = 0.46 F S$^{-1}$ cm$^{-2}$ respectively.

## Introduction

There is a continuous interest in developing multifunctional materials for energy conversion and storage systems that are also able to exhibit additional mechanical, thermal or optical properties such as flexibility and transparency. The envisaged applications for these multifunctional materials include power supply,[1] displays,[2] smart windows,[3,4] transistors,[5] light-emitting devices,[6] touch sensors[7,8] and photovoltaics[9]. A key class of active materials in these efforts are nanocarbons, such as graphene and carbon nanotubes (CNTs). Their high electrochemical stability and electrical conductivity make them particularly suitable for operation in supercapacitors with high specific power, long cycling life and fast charge-discharge rate.[10,11] Moreover, the ability to process graphene and CNTs as porous and mechanically stable electrodes/current collectors makes them attractive for a wide range of devices with augmented properties relative to standard monofunctional systems, such as flexible[12] and structural supercapacitors[13,14] and flexible solar cells[15,16]. Is has been also demonstrated that, when those nanocarbons are processed as thin, percolating networks, their optical absorption is dramatically reduced, which makes them suitable for various application in transparent electronics, photonics, and optoelectronics,[17–19] among others.

In the last years, substantial efforts to use nanocarbon-based networks as active electrode material in transparent electric double-layer capacitors (EDLCs) have been made. The predominant strategy is to prepare a dispersion of high purity single walled carbon nanotubes (SWNT), graphene or a combination, and then deposit a thin layer onto a transparent substrate, followed by introduction of the electrolyte and a second thin electrode layer. This procedure has led to multifunctional devices with a wide range of properties; e.g. power densities of 0.9 – 120 kW kg$^{-1}$, energy densities of 0.9 – 20 Wh kg$^{-1}$ and optical transmittance of 50 – 80% at 550 nm.[20–29] The upper range of these EDLCs has comparable properties to flexible, transparent supercapacitors based on pseudocapacitive conductive polymers and battery-type metal oxides, albeit with superior cyclability.[30–35]

The first requirement to obtain good-quality thin electrodes relies on achieving excellent nanocarbon dispersions. However, dispersing nanocarbons typically requires their purification, size-selection, shortening, or chemical functionalization, which not only introduces multiple processing stages, but can also reduce the figure-of-merit for transparent conductors (TCs) by transitioning from bulk to percolating behaviours.[36,37] For this reason, it is crucial to explore other methods to assemble nanocarbon based transparent electrodes for ELDCs, particularly those that avoid nanocarbon dispersions. An alternative approach is deposition of CNTs directly from the gas-phase as they are grown by floating catalyst chemical vapour deposition (CVD). This method has consistently led to superior TCs with a sheet resistance of up to 41 – 820 Ω cm$^{-2}$ at 90% of optical transmittance, using a combination of advanced molecular control and by keeping a low concentration of CNTs in the gas-phase (≈5 x 10$^5$ cm$^{-3}$).[27,28,38–43] Electrodes made up of CNT fibres produced by direct spinning of the aerogel have also been successfully used in various high-power supercapacitors with large bending flexibility and toughness,[12,44,45] and more recently incorporated into structural composites.[13] In such process the CNTs grown floating in the gas are at much higher concentration (1 x 10$^9$ cm$^{-3}$),[46] thus forming a continuous network that can be directly process as macroscopic materials.

Motivated by these results, this work reports the fabrication of highly-flexible and transparent supercapacitors with ultra-high power density by directly depositing thin CNT fibre electrodes from the gas-phase onto an ionic liquid-based polymer electrolyte (PE) membrane. The simple assembling procedure consists in direct integration of a transparent CNT fibre film from a floating catalyst CVD reactor onto both sides of a PE membrane. Then, two electrical terminals are directly connected to the sandwich-like device structure followed by stacking and laminating in a transparent plastic pouch cell. The excellent electrical conductivity of CNT fibres precludes the use of current collector and leads to devices combining high optical transparency (*ca.*70%), high energy density of 10 Wh kg$^{-1}$, and record-high power density of 1370 kW kg$^{-1}$. These values represent an improvement of 1-3 orders of magnitude relative to prior-art transparent EDLCs based on nanocarbons.

## Results and discussion

Our general method to produce CNT fibre electrodes involves direct spinning of a CNT aerogel from the gas phase during growth of the CNTs by floating catalyst chemical vapor deposition (FC-CVD) (see Supporting Information). At the exit of the CVD reactor, the low-density translucent CNT aerogel is wound onto a spool, which can optionally have a substrate (*Figure 1a*). Conventional electrodes are built up by winding multiple single filaments over prolonged time (*Figure 1b*), whereas the transparent material is produced by collecting a single aerogel filament onto a transparent substrate, such as mica or a polymer sheet (*Figure 1c*). In a previous work, we studied the electrochemical properties of CNT fibre electrodes in an ionic liquid electrolyte which showed that specific power ($P$) values scale with the reciprocal of electrode areal density ($m$), commonly known as mass loading ($P = 3.34/m$, with power normalized by mass per unit area in mg cm$^{-2}$ of active material in both electrodes ).[12] This behavior reflects the linear increase in discharge time with increasing mass loading which is common in highly-conducting EDLC electrodes where capacitance is not limited by mass transport.[47] Very importantly, such analysis predicts that utilization of single filament electrodes (m ≈ 1 µg cm$^{-2}$) should result in transparent devices with power densities exceeding 1 MW kg$^{-1}$.

First, we focus briefly on the electrical and optical properties of CNT fibre transparent conductors (TCs). They are produced by depositing single CNT fibre aerogel filaments onto a transparent support, as shown in *Figure 2a*. The resulting material consists of a continuous porous network of bundles (*Figure 2b*) comprising of highly graphitized few-layer multiwalled CNTs (*Figure 2c*) with estimated length of 1 mm.[46] The areal density of these samples can be conveniently varied by overlapping more layers at the point of collection. To facilitate electrical measurements, the samples were directly deposited onto substrates with four equidistant pre-patterned gold electrodes (see Supporting Information).

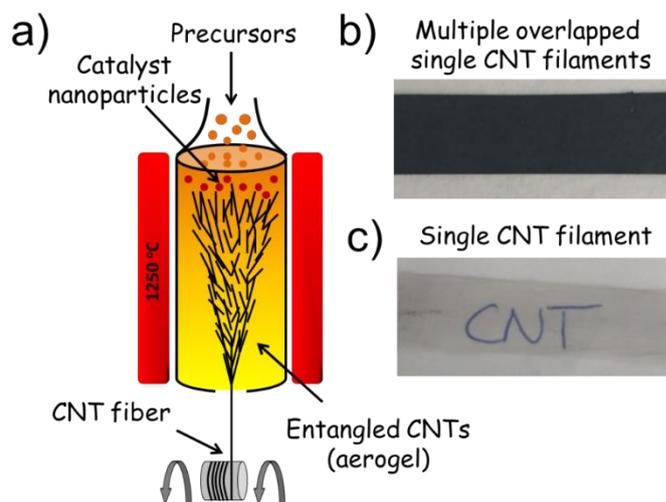

**Fig. 1.** a) Schematic of the fabrication of electrodes by collection of a CNT aerogel directly from the gas-phase during growth by floating catalyst chemical vapour deposition. b) Electrode thickness is built by collection of multiple aerogel layers on the same spool. c) Single CNT aerogel filament can be deposited onto a transparent supporting substrate to make a transparent conducting electrode.

*Figure 2d* depicts a plot of optical transmittance (T) at 550 nm as a function of sheet resistance (R$_s$) for the CNT fibre sheet produced in this work (the optical transmittance spectra for 1-5 layers of CNT fibres are shown in *Figure 1Sa, Supporting Information*). The results show that the material is a good TC, combining high transparency and relatively low R$_s$ reaching, for example, 462 Ohms sq$^{-1}$ at 90% of transmittance. A common electrical figure of merit (FoM$_e$) for TCs is the ratio of DC conductivity and optical conductivity ($\sigma_{DC}/\sigma_{op}$) included in the following equation:

$$T = (1 + \frac{188.5}{R_s}\frac{\sigma_{op}}{\sigma_{DC}})^{-2}.$$

(1)

Fitting the experimental values of T and R$_s$ using Equation 1 (red dashed line) translates into a FoM$_e$ of 2.7. This value is similar to those reported for related floating catalyst (FC)-CVD processes in the absence of dopants [42,48–51], and comparable to transparent nanocarbon electrodes based on wet-processed SWCNTs,[52,53] graphene,[54,55] rGO[56,57] and their hybrids[58–60] (see *Figure 1Sb and Table1s, Supporting Information*). The plot of (T$^{(-0.5)}$ – 1) as a function of R$_s$ in *Figure 2e* confirms that CNT fibres demonstrate bulk-like conductive behavior in whole optical transmittance range. Unlike in other nanostructured transparent conductor electrodes of small aspect ratio conductors,[36,37] reducing CNT fibre thickness does not produce transition below the percolating regime. The extraordinarily high aspect ratio (10$^6$) of the constituent CNTs in this process implies that the transition from bulk to percolation behavior is below practical thickness attainable in an electrode using this preparation method. Additionally, note that the actual fibre drawing process requires mechanical percolation in the material to sustain the associated stresses when the fibre is withdrawn from the reactor,[61] thus implying that samples produced continuously are inherently in the bulk regime.

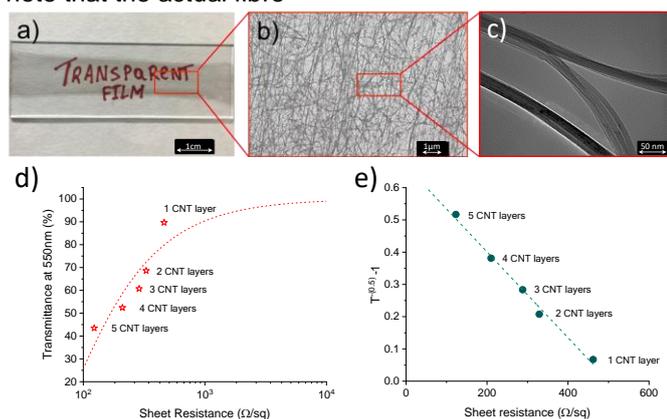

**Fig. 2.** a) Photo of single layer of CNT fibre, b) SEM and c) TEM images demonstrating morphology and microstructure of CNTs network and their highly graphitic structure. d) The plot of transmittance at 550nm as a function of sheet resistance and e) corresponding plot of $T^{(-0.5)} - 1$ as a function of $R_s$, following nearly linear dependence and confirming bulk-like conductive behaviour of the CNT fibres in all range of transmittance.

From equation fitting in *Figure 2e* and assuming optical conductivity of $2.0 \times 10^4$ S m$^{-1}$[62,63], DC conductivity comes out as $5.4 \times 10^4$ S m$^{-1}$, which is in the range of longitudinal conductivities measured for similar CNT fibre materials and higher than that of high-quality solution-processed CNT networks at $3 \times 10^4$ S m$^{-1}$.[37] A small degree of CNT alignment might contribute to such high values of conductivity, but in general the requirement of transparency implies that the network is largely misoriented (*Figure 2b*). In this FC-CVD process the conducting network morphology is reminiscent of the aerogel structure. However, although aerogel volumetric density can be controlled by adjusting gas flow rate in the reactor or the spinning/winding rate, this mainly affects areal density of the TC but not its FoM$_e$ (see *Figure 2S, Supporting Information*).[64] These properties, combined with the evidence of a high specific surface area (~250 m$^2$ g$^{-1}$) and mesoporous structure in bulk CNT fibre[65], makes transparent CNT fibre electrodes promising material for application in EDLCs. As a consequence of CNT aggregation, necessary to form a conducting percolating network, specific surface area is lower than theoretical values for individualised CNTs,[66] but nevertheless much higher than for metallic conductors.

The electrochemical properties of single filament CNT fibre were evaluated by assembling a symmetric coin-shaped EDLC device. A flexible polymer electrolyte membrane, consisting of 70 wt.% of N-methyl-N-butylpyrrolidinium bis(trifluoromethanesulfonyl)imide (Pyr$_{14}$TFSI) ionic liquid and 30 wt.% of poly(vinylidene fluoride-co-hexafluoropropylene) (PVDF-co-HFP) polymer, was used as separator and electrolyte source. The mass loading of CNT fibre was 1.75 ± 0.05 µg cm$^{-2}$, determined from several measurements of fibre's gravimetric linear density (see Supporting Information). Accordingly, the total mass loading of active material in the full device was 3.50 ± 0.05 µg cm$^{-2}$.

The results of galvanostatic charge-discharge (CD) tests at different current densities in a two-electrode Swagelok cell are presented in *Figure 3*. A comparison of the CD profiles of the EDLC devices based on single-filament CNT electrodes (3.5 µg cm$^{-2}$) and conventional high mass loading CNT fibres (1.27 mg cm$^{-2}$) is shown in *Figure 3a*. The triangular shape of the charge-discharge capacity with a constant slope along the whole dis(charge) voltage undoubtedly demonstrates the electrostatic nature of the charge storing mechanism. In *Figure 3a* both curves exhibit similar triangular shape with high coulombic efficiencies ($\mu = Q_{discharge}/Q_{charge} = (I_d \ast t_d)/(I_c \ast t_c)$) (>95%), demonstrating reversible capacitive behavior of both devices. Noticeably, similar values of the ohmic drop indicate negligible influence of electrode's thickness on equivalent series resistance (ESR) of EDLCs. Moreover, ESR is in the range of values reported for similar electrochemical system[12], and it does not substantially vary at different currents applied (see *Figure 3S, Supporting Information*). Moreover, the discharge profiles in *Figure 3a* exhibit fairly similar slopes which indicates that the thickness of the electrode does not significantly affect the specific capacitance (C ~ 1/slope) of the devices. In fact, specific capacitances of single filament EDLC are slightly higher than those for the high mass loading EDLC in the whole range of current densities (*Figure 3b*), achieving maximum values of 31.2 and 27.5 F g$^{-1}$ at 1 mA cm$^{-2}$, respectively. This small difference can be attributed to a more effective impregnation of the polymer electrolyte into the ultrathin film of single filament electrode maximizing the material utilization.

For reference, specific energy (E$_{real}$) and power (P$_{real}$) values were calculated from the discharge profiles and compared in a Ragone plot in *Figure 3c*. The maximum value of E$_{real}$ achieved at 1 mA cm$^{-2}$ for the single filament EDLC was 12.3 Wh kg$^{-1}$ (84.6 nWh cm$^{-2}$), whilst P$_{real}$ reached the highest value of 1579 kW kg$^{-1}$ at 10 mA cm$^{-2}$ corresponding to areal power density of 10.86 mW cm$^{-2}$. Remarkably, the value of specific power at 5 mA cm$^{-2}$ was found to be still ~1.0 MW kg$^{-1}$.

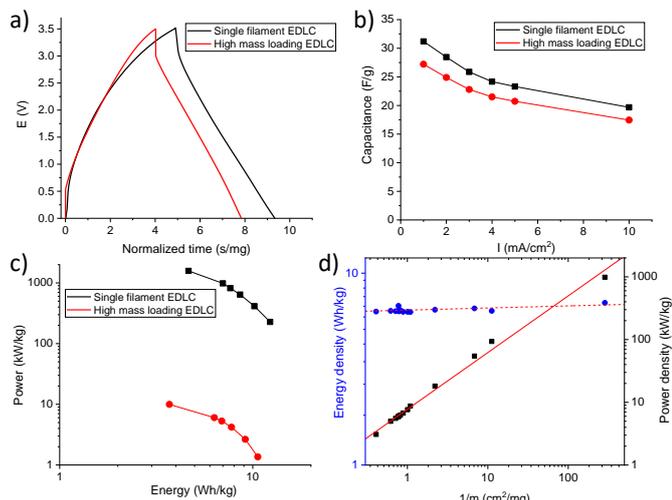

**Fig. 3.** Comparison of single-filament (3.5 µg cm$^{-2}$) and high mass loading (1.27 mg cm$^{-2}$) CNT fibre-based EDLCs: a) CD profiles obtained at 5 mA cm$^{-2}$, b) specific capacitance (C$_s$) at different current densities and c) Ragone plots. d) Real energy and power densities of symmetric supercapacitors at different total mass loading of CNT fibres, calculated from CD measurements performed at 3.5V operating voltage and 5 mA cm$^{-2}$ current density.

To further investigate the effect of electrode thickness on the electrochemical performance, the specific power of samples with fixed area but different electrode mass loading (i.e. thickness) ranging from 3.5 µg cm$^{-2}$ to 2.40 mg cm$^{-2}$ is compared. The plots of real energy and power densities against the reciprocal of electrode mass loading (considering the active mass of two electrodes), are presented in *Figure 3d*. As anticipated, the specific energy remains nearly constant, whereas the power density shows a linear dependence on the inverse of mass loading (1/m). This confirms that the CNT fibre electrodes behave as a bulk conductive material in EDLC device, which is a key feature enabling to achieve the extremely high power densities.

With these results in mind, we explored the behavior of free-standing transparent EDLC devices composed of CNT fibre electrodes and the polymer electrolyte encapsulated into a transparent plastic pouch cell. The high electrical conductivity of CNT fibres associated to its unique structure enables its dual role as active material and current collector, using a conducing frame made with simple aluminum foil for external connections (*Figure 4a*). The optical transmittance values of the device and its individual components (PE membrane and single filament CNT electrodes) are depicted in *Figure 4b*. The high transmittance of the PE membrane (98%) and of each single filament CNT fibre (~90%) leads to a device with *ca.* 70% optical transmittance at a wavelength of 550 nm.

Charge-discharge profiles obtained for a flexible and transparent free-standing EDLC compared with a similar EDLC assembled in a typical Swagelok cell are included in *Figure 4c*. This comparison is relevant since results in two-electrode Swagelok or coin cells are often considered to represent an upper limit for the devices. The maximum values of real power and energy densities for the transparent free-standing device are as high as 1370 kW kg$^{-1}$ and 10 Wh kg$^{-1}$, respectively, with specific capacitance (C$_s$) of 29 F g$^{-1}$. These values only differ slightly from those obtained using Swagelok cell configuration mainly due to the larger equivalent series resistance (ESR) (50 Ohms vs 40 Ohms) exhibited by the transparent free-standing device. The higher ESR was also corroborated by EIS measurements showing a larger semicircle at high frequencies observed in the Nyquist plot of the transparent device (*Figure 4d*). We attribute this difference to the higher pressure and resultant better compaction of EDLC sandwich structure in the Swagelok cell producing improved electrical contact between the electrodes and current collector. The maximum energy ($E_{max} = CV^2/2$) and power ($P_{max} = V^2/(4*ESR)$) densities were also calculated for comparative purposes and found to be 12.3 Wh kg$^{-1}$ and 7290 kW kg$^{-1}$, respectively.

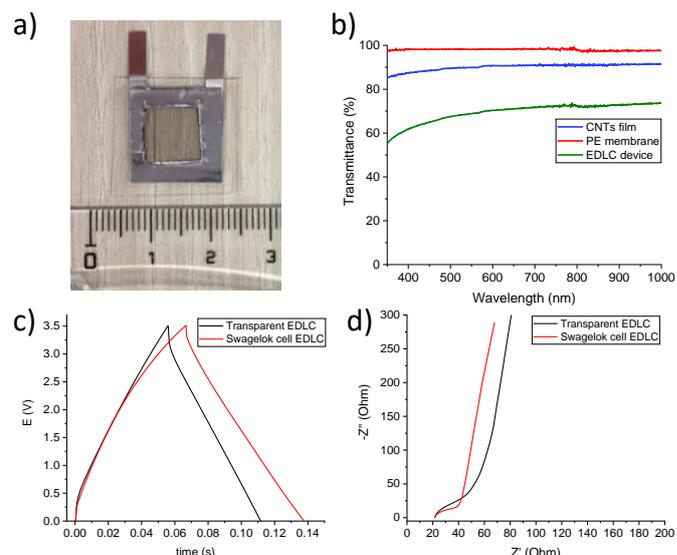

**Fig. 4.** a) Photograph of free-standing transparent EDLC device, b) Optical transmittance of CNT fibre/PE/CNT fibre composite, PE membrane and single layer of CNT fibre, c) CD profiles at 2 mA cm$^{-2}$ and d) Nyquist plots comparing performance of single-filament (3.5 µg cm$^{-2}$) CNT fibre-based EDLCs assembled in a Swagelok cell and laminated into a transparent free-standing flexible device.

To compare the capacitive performance of transparent CNT fibre films to the literature data, we have calculated capacitive figure of merit (FoM$_c$ = C$_V$/σ$_{op}$) using the following equation:

$$T = (1 + \frac{188.5\sigma_{op}}{C_V} c_A)^{-2}.$$

(2)

Here C$_V$ and C$_A$ are the volumetric and areal capacitances, respectively. Since gravimetric capacitance was found to be independent of CNT fibre mass loading (see *Figure 3*) and considering a value of C$_s$ = 78 F g$^{-1}$ per electrode,[65] FoM$_c$ for single CNT fibre layer (1.75 µg cm$^{-2}$) corresponds to 0.46 F S$^{-1}$ cm$^{-2}$. This value is higher than those reported previously for transparent electrodes based on SWCNTs

(0.32 F S$^{-1}$ cm$^{-2}$), confirming excellent multifunctional properties of single-filament CNT fibre electrodes.[37]

As-fabricated free-standing transparent EDLC device was subjected to 20 000 charge–discharge cycles in a voltage window of 3.5 V at 5 mA cm$^{-2}$ to study its long-term cycling performance. The values of specific capacitance and energy density were not significantly affected during cycling. *Figure 5a* shows that capacitance retention after cycling was near 94%, with a drop in energy density of less than 4% after 20 000 cycles (and coulombic efficiency values remaining constant at ~99% as depicted in *Figure 4S*) confirming the excellent long-term electrochemical performance of the system. These results are particularly relevant considering the high voltage used during charging, and the fact that the nominal 5 mA cm$^{-2}$ current density applied corresponds to a very high effective in-plane current density when normalized by CNT fibre cross sectional area (~ 1.1 ×10$^7$ Am$^{-2}$).

To explore a potential application of the assembled device in flexible electronics, we have additionally studied the electrochemical performance of the transparent EDLC under flexural stresses by subjecting it to 100 cycles of 180° bending. As it is shown in *Figure 5b,* the CD profiles before and after bending have identical line shapes, indicating that repeating bending of the EDLC does not induce any deterioration of its electrochemical performance. Such robustness origins in the intrinsic toughness of the CNT fibre electrodes and the formation of a large ductile interface upon infiltration of the PE.

*Figure 6* summarizes the performance of transparent and flexible supercapacitors reported in literature (cross-sectional plots included in *Supporting Information, Figure 5S*). Most of the studies report E$_{max}$ and P$_{max}$, which can significantly alter from realistic available energy and power of a system. However, the lack of real energy and power values in the literature leads to the necessity of representing E$_{max}$ and P$_{max}$ for accurate comparison of the performances. Transparent and flexible device based on single layer CNT fibre electrodes and IL-based PE exhibits good transmittance combined with outstanding electrochemical properties among the highest in state-of-the-art. In fact, the power density of the EDLC based on CNT fibre sheets has substantially outperformed previously reported transparent supercapacitors based on graphene,[20,23,29] CNTs,[24,25,27] and rGO[21] by 1 to 4 orders of magnitude, while also showing large energy density. It is worth mentioning that even real available specific power of our transparent EDLC significantly surpasses P$_{max}$ values from previous studies. Interestingly, comparison with the literature also reveals orders-of-magnitude larger current densities achieved for the present CNT fibre-based EDLC (10$^{-3}$-10$^{-2}$ A/cm$^2$ vs 10$^{-7}$-10$^{-4}$ A/cm$^2$) leading to much faster discharge rates (t$_{dis}$ <0.2s vs 10-100s). This additionally indicates exceptional conductive performance of thin CNT fibre films. Therefore, the combination of good optical transparency, high flexibility, and excellent energy storage capabilities makes EDLC based on CNT fibres and PE membrane a highly promising multifunctional energy storage system.

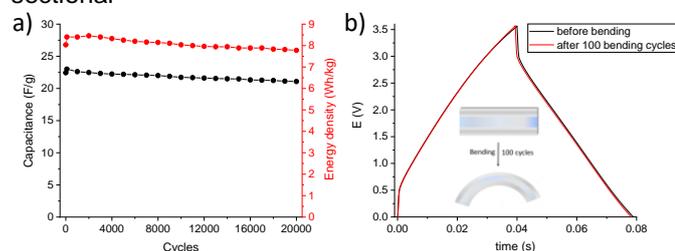

**Fig. 5.** a) CD cycling stability test of the transparent free-standing single filament EDLC at 5 mA cm$^{-2}$ and operating voltage of 3.5V showing retention of specific capacitance (C$_s$) and energy density over 20,000 cycles and b) CD profiles at 5 mA cm$^{-2}$ demonstrating reproducible electrochemical performance after 100 bending cycles.

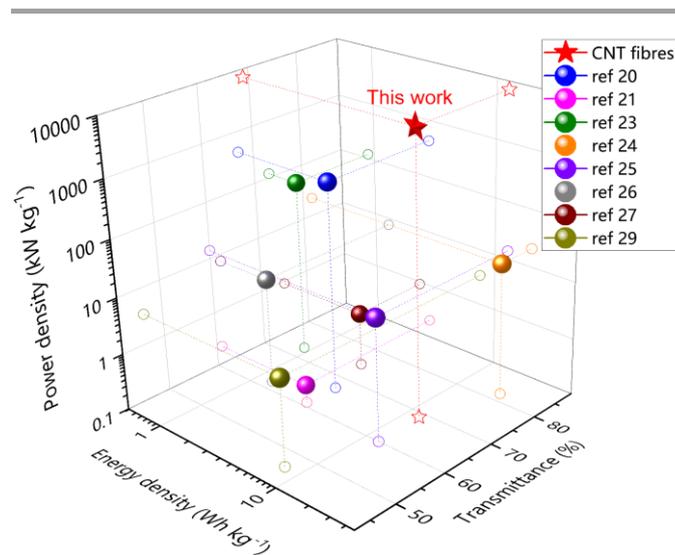

**Fig. 6.** 3D plot comparing E$_{max}$, P$_{max}$ and optical transmittance at 550 nm for full-cell transparent supercapacitors presented in this work and reported in literature.[20,21,23–27,29] Detailed values are included in Table S2 (Supporting Information).

## Conclusions

In summary, we have produced a multifunctional transparent and flexible EDLC based on a polymer electrolyte and ultra-thin transparent electrodes of CNT fibres directly drawn as an aerogel from the gas-phase during CNT growth. When used as transparent conductor, interconnected network of CNTs in as-spun fibres provides a FoM$_e$ of 2.7, reaching for example 566 Ohm sq$^{-1}$ Rs at 90% transparency. These properties stem from the exceptionally long length of the constituent CNTs, making the electrodes behave as bulk conductors, thus showing both higher conductivity and transparency compared to solution-processed percolating networks of CNTs. The FoM$_e$ seems to be fixed for a given CNT fibre composition, with changes in aerogel morphology affecting simply areal density. In view of these results, work is in progress to study the effect of CNT type and bundle dimensions as routes for further performance improvement.

The flexible and transparent supercapacitor devices were found to exhibit an optimal transmittance of *ca.*- 70% in combination with an outstanding electrochemical performance including a power density of 1370 kW kg$^{-1}$, the largest achieved in transparent EDLC configuration. A linear dependence of power density on 1/electrode mass loading was obtained experimentally for these devices. It confirms the efficiency of the assembly process, while also reflecting the highly conducting nature of the CNT fibre electrodes. The transparent EDLC device also show excellent cycling stability over 20,000 cycles at 3.5 V and no changes in electrochemical performance after 100 bending cycles at 180°.

The current capacitive figure of merit for CNT fibre film is 0.46 F S$^{-1}$ cm$^{-2}$, which is among the largest FOM$_c$ values for pristine nanocarbon electrodes. This is compensated by the enormous simplicity of the synthetic process used here. Based on previous experience in fibre spinning and fabrication of large-area all-solid supercapacitors, we envisage that transparent supercapacitors based on CNT fibres could be produced continuously at rates as high as 100 m/min. More importantly, further improvements in capacitance by introducing pseudocapacitve reactions are easily within hand.

## Conflicts of interest

There are no conflicts to declare.

## Acknowledgements

Financial support is acknowledged from the European Union Seventh Framework Program under grant agreement 678565 (ERC-STEM), from MINECO (RyC-2014-15115, Spain) and CAM MAD2D project (S2013/MIT-3007). R.D.C. acknowledges the Spanish MINECO for the Ramón y Cajal program (RYC-2016-20891) and HYNANOSC (RTI2018-099504-A-C22). R.D.C and J.J.V. also acknowledges the FOTOART-CM project funded by Madrid region under programme P2018/NMT-4367. J. F.-C. acknowledges the Marie Skłodowska-Curie Individual Fellowships (H2020-MSCA-IF-2017). M.V. acknowledges the Madrid Regional Government (program "Atracción de Talento Investigador", 2017-T2/IND-5568). RM thanks the Spanish Ministry of Science, Innovation and Universities through the SUSBAT project (RTI2018-101049-B-I00) (MINECO/FEDER, UE) and the Excellence Network E3TECH (CTQ2017-90659-REDT (MINECO, Spain).

## References

1. J. Luo, W. Tang, F. R. Fan, C. Liu, Y. Pang, G. Cao and Z. L. Wang, *ACS Nano*, 2016, **10**, 8078–8086.
2. M. Zhu, Z. Wang, H. Li, Y. Xiong, Z. Liu, Z. Tang, Y. Huang, A. L. Rogach and C. Zhi, *Energy Environ. Sci.*, 2018, **11**, 2414–2422.
3. P. Yang, P. Sun, Z. Chai, L. Huang, X. Cai, S. Tan, J. Song and W. Mai, *Angew. Chemie Int. Ed.*, 2014, **53**, 11935–11939.
4. K. Wang, H. Wu, Y. Meng, Y. Zhang and Z. Wei, *Energy Environ. Sci.*, 2012, **5**, 8384–8389.
5. D. W. Kim, S.-Y. Min, Y. Lee and U. Jeong, *ACS Nano*, 2020, **14**, 907–918.
6. S. Ju, J. Li, J. Liu, P.-C. Chen, Y. Ha, F. Ishikawa, H. Chang, C. Zhou, A. Facchetti and D. B. Janes, *Nano Lett.*, 2008, **8**, 997–1004.
7. M. R. Kulkarni, R. A. John, M. Rajput, N. Tiwari, N. Yantara, A. C. Nguyen and N. Mathews, *ACS Appl. Mater. Interfaces*, 2017, **9**, 15015–15021.
8. B. Meng, W. Tang, Z. Too, X. Zhang, M. Han, W. Liu and H. Zhang, *Energy Environ. Sci.*, 2013, **6**, 3235–3240.
9. A. Cannavale, G. E. Eperon, P. Cossari, A. Abate, H. J. Snaith and G. Gigli, *Energy Environ. Sci.*, 2015, **8**, 1578–1584.
10. D. S. Su and R. Schlögl, *ChemSusChem Chem. Sustain. Energy Mater.*, 2010, **3**, 136–168.
11. A. Ghosh and Y. H. Lee, *ChemSusChem*, 2012, **5**, 480–499.
12. E. Senokos, V. Reguero, L. Cabana, J. Palma, R. Marcilla and J. J. Vilatela, *Adv. Mater. Technol.*, 2017, **2**, 1600290.
13. E. Senokos, Y. Ou, J. J. Torres, F. Sket, C. González, R. Marcilla and J. J. Vilatela, *Sci. Rep.*, 2018, **8**, 3407.
14. H. Qian, A. R. Kucernak, E. S. Greenhalgh, A. Bismarck and M. S. P. Shaffer, *ACS Appl. Mater. Interfaces*, 2013, **5**, 6113.
15. A. Monreal-Bernal, J. J. Vilatela and R. D. Costa, *Carbon N. Y.*, 2019, **141**, 488–496.
16. Z. Liu, J. Li and F. Yan, *Adv. Mater.*, 2013, **25**, 4296–4301.
17. C. J. Zhang and V. Nicolosi, *Energy Storage Mater.*, 2019, **16**, 102–125.
18. S. B. Yang, B.-S. Kong, D.-H. Jung, Y.-K. Baek, C.-S. Han, S.-K. Oh and H.-T. Jung, *Nanoscale*, 2011, **3**, 1361–1373.
19. L. Yu, C. Shearer and J. Shapter, *Chem. Rev.*, 2016, **116**, 13413–13453.
20. Y. Gao, Y. S. Zhou, W. Xiong, L. J. Jiang, M. Mahjouri-Samani, P. Thirugnanam, X. Huang, M. M. Wang, L. Jiang and Y. F. Lu, *APL Mater.*, 2013, **1**, 12101.
21. T. Aytug, M. S. Rager, W. Higgins, F. G. Brown, G. M. Veith, C. M. Rouleau, H. Wang, Z. D. Hood, S. M. Mahurin and R. T. Mayes, *ACS Appl. Mater. Interfaces*, 2018, **10**, 11008–11017.
22. E. P. Gilshteyn, T. Kallio, P. Kanninen, E. O. Fedorovskaya, A. S. Anisimov and A. G. Nasibulin, *RSC Adv.*, 2016, **6**, 93915–93921.
23. P. Xu, J. Kang, J.-B. Choi, J. Suhr, J. Yu, F. Li, J.-H. Byun, B.-S. Kim and T.-W. Chou, *ACS Nano*, 2014, **8**, 9437–9445.
24. R. Yuksel, Z. Sarioba, A. Cirpan, P. Hiralal and H. E. Unalan, *ACS Appl. Mater. Interfaces*, 2014, **6**, 15434–15439.

25. Z. Niu, W. Zhou, J. Chen, G. Feng, H. Li, Y. Hu, W. Ma, H. Dong, J. Li and S. Xie, *Small*, 2013, **9**, 518–524.
26. P.-C. Chen, G. Shen, S. Sukcharoenchoke and C. Zhou, *Appl. Phys. Lett.*, 2009, **94**, 43113.
27. T. Chen, H. Peng, M. Durstock and L. Dai, *Sci. Rep.*, 2014, **4**, 3612.
28. S. Chun, W. Son, G. Lee, S. H. Kim, J. W. Park, S. J. Kim, C. Pang and C. Choi, *ACS Appl. Mater. Interfaces*, 2019, **11**, 9301–9308.
29. N. Li, G. Yang, Y. Sun, H. Song, H. Cui, G. Yang and C. Wang, *Nano Lett.*, 2015, **15**, 3195–3203.
30. J. Liu, G. Shen, S. Zhao, X. He, C. Zhang, T. Jiang, J. Jiang and B. Chen, *J. Mater. Chem. A*, 2019, **7**, 8184–8193.
31. X. Y. Liu, Y. Q. Gao and G. W. Yang, *Nanoscale*, 2016, **8**, 4227–4235.
32. T. Cheng, Y.-Z. Zhang, J.-D. Zhang, W.-Y. Lai and W. Huang, *J. Mater. Chem. A*, 2016, **4**, 10493–10499.
33. S. B. Singh, T. I. Singh, N. H. Kim and J. H. Lee, *J. Mater. Chem. A*, 2019, **7**, 10672–10683.
34. H. Sheng, X. Zhang, Y. Ma, P. Wang, J. Zhou, Q. Su, W. Lan, E. Xie and C. J. Zhang, *ACS Appl. Mater. Interfaces*, 2019, **11**, 8992–9001.
35. C. J. Zhang, T. M. Higgins, S.-H. Park, S. E. O'Brien, D. Long, J. N. Coleman and V. Nicolosi, *Nano Energy*, 2016, **28**, 495–505.
36. S. De and J. N. Coleman, *MRS Bull.*, 2011, **36**, 774–781.
37. P. J. King, T. M. Higgins, S. De, N. Nicoloso and J. N. Coleman, *ACS Nano*, 2012, **6**, 1732–1741.
38. Q. Cheng, J. Tang, J. Ma, H. Zhang, N. Shinya and L.-C. Qin, *Phys. Chem. Chem. Phys.*, 2011, **13**, 17615–24.
39. P. Chen, H. Chen, J. Qiu and C. Zhou, *Nano Res.*, 2010, **3**, 594–603.
40. T. Carlson, D. Ordéus, M. Wysocki and L. E. Asp, *Compos. Sci. Technol.*, 2010, **70**, 1135–1140.
41. A. Balducci, S. S. Jeong, G. T. Kim, S. Passerini, M. Winter, M. Schmuck, G. B. Appetecchi, R. Marcilla, D. Mecerreyes, V. Barsukov, V. Khomenko, I. Cantero, I. De Meatza, M. Holzapfel and N. Tran, *J. Power Sources*, 2011, **196**, 9719–9730.
42. S. Ahmad, E.-X. Ding, Q. Zhang, H. Jiang, J. Sainio, M. Tavakkoli, A. Hussain, Y. Liao and E. I. Kauppinen, *Chem. Eng. J.*, 2019, **378**, 122010.
43. K. Mustonen, On the limit of single-walled carbon nanotube random network conductivity, 2015.
44. A. Pendashteh, E. Senokos, J. Palma, M. Anderson, J. J. Vilatela and R. Marcilla, *J. Power Sources*, 2017, **372**, 64–73.
45. D. Iglesias, E. Senokos, B. Alemán, L. Cabana, C. Navío, R. Marcilla, M. Prato, J. J. Vilatela and S. Marchesan, *ACS Appl. Mater. Interfaces*, 2018, **10**, 5760–5770.
46. V. Reguero, B. Alemán, B. Mas and J. J. Vilatela, *Chem. Mater.*, 2014, **26**, 3550–3557.
47. D. Dunn and J. Newman, *J. Electrochem. Soc.*, 2000, **147**, 820–830.
48. E.-X. Ding, A. Hussain, S. Ahmad, Q. Zhang, Y. Liao, H. Jiang and E. I. Kauppinen, *Nano Res.*, 2019, 1–9.
49. A. Hussain, Y. Liao, Q. Zhang, E.-X. Ding, P. Laiho, S. Ahmad, N. Wei, Y. Tian, H. Jiang and E. I. Kauppinen, *Nanoscale*, 2018, **10**, 9752–9759.
50. A. Kaskela, A. G. Nasibulin, M. Y. Timmermans, B. Aitchison, A. Papadimitratos, Y. Tian, Z. Zhu, H. Jiang, D. P. Brown and A. Zakhidov, *Nano Lett.*, 2010, **10**, 4349–4355.
51. A. G. Nasibulin, A. Kaskela, K. Mustonen, A. S. Anisimov, V. Ruiz, S. Kivisto, S. Rackauskas, M. Y. Timmermans, M. Pudas and B. Aitchison, *ACS Nano*, 2011, **5**, 3214–3221.
52. A. P. Tsapenko, A. E. Goldt, E. Shulga, Z. I. Popov, K. I. Maslakov, A. S. Anisimov, P. B. Sorokin and A. G. Nasibulin, *Carbon N. Y.*, 2018, **130**, 448–457.
53. Q. Zhang, N. Wei, P. Laiho and E. I. Kauppinen, *Top. Curr. Chem.*, 2017, **375**, 90.
54. X. Li, G. Zhang, X. Bai, X. Sun, X. Wang, E. Wang and H. Dai, *Nat. Nanotechnol.*, 2008, **3**, 538.
55. K. S. Kim, Y. Zhao, H. Jang, S. Y. Lee, J. M. Kim, K. S. Kim, J.-H. Ahn, P. Kim, J.-Y. Choi and B. H. Hong, *Nature*, 2009, **457**, 706.
56. H. A. Becerril, J. Mao, Z. Liu, R. M. Stoltenberg, Z. Bao and Y. Chen, *ACS Nano*, 2008, **2**, 463–470.
57. J. Zhao, S. Pei, W. Ren, L. Gao and H.-M. Cheng, *ACS Nano*, 2010, **4**, 5245–5252.
58. Q. Zheng, B. Zhang, X. Lin, X. Shen, N. Yousefi, Z.-D. Huang, Z. Li and J.-K. Kim, *J. Mater. Chem.*, 2012, **22**, 25072–25082.
59. Y. Liao, K. Mustonen, S. Tulic, V. Skakalova, S. A. Khan, P. Laiho, Q. Zhang, C. Li, M. R. A. Monazam and J. Kotakoski, *ACS Nano*, 2019, **13**, 11522–11529.
60. S. De, P. J. King, M. Lotya, A. O'Neill, E. M. Doherty, Y. Hernandez, G. S. Duesberg and J. N. Coleman, *Small*, 2010, **6**, 458–464.
61. B. Alemán, V. Reguero, B. Mas and J. J. Vilatela, *ACS Nano*, 2015, **9**, 7392–8.
62. D. S. Hecht, A. M. Heintz, R. Lee, L. Hu, B. Moore, C. Cucksey and S. Risser, *Nanotechnology*, 2011, **22**, 75201.
63. B. Ruzicka, L. Degiorgi, R. Gaal, L. Thien-Nga, R. Bacsa, J.-P. Salvetat and L. Forró, *Phys. Rev. B*, 2000, **61**, R2468–R2471.
64. I. S. Fraser, M. S. Motta, R. K. Schmidt and A. H. Windle, *Sci. Technol. Adv. Mater.*, 2010, 11, 45004.
65. E. Senokos, V. Reguero, J. Palma, J. J. Vilatela and R. Marcilla, *Nanoscale*, 2016, **8**, 3620–3628.
66. A. Peigney, C. Laurent, E. Flahaut, R. R. Bacsa and A. Rousset, *Carbon N. Y.*, 2001, **39**, 507–514.